\documentclass[twocolumn,aps,pra,amsmath,floatfix]{revtex4}
\usepackage{graphicx}
\begin{document}

\title{Coulomb correlations do not fill the $e'_g$ hole pockets in 
       Na$_{0.3}$CoO$_2$} 
\author{A. Liebsch$^1$ and H. Ishida$^2$ } 
\affiliation{$^1$Institut f\"ur Festk\"orperforschung, Forschungszentrum
             J\"ulich, 52425 J\"ulich, Germany\\
             $^2$College of Humanities and Sciences, Nihon University,
             and CREST JST, Tokyo 156, Japan}
\date{\today }

\begin{abstract}
There exists presently considerable debate over the question whether
local Coulomb interactions can explain the absence of the small $e'_g$ 
Fermi surface hole pockets in photoemission studies of Na$_{0.3}$CoO$_2$.
By comparing dynamical mean field results for different single particle
Hamiltonians and exact diagonalization as well as quantum Monte Carlo
treatments, we show that, for realistic values of the Coulomb energy $U$ 
and Hund exchange $J$, the $e'_g$ pockets can be slightly enhanced or 
reduced compared to band structure predictions, but they do not disappear.
\end{abstract}
\maketitle

The Fermi surface of a material is one of its most fundamental properties.
Usually, it can be understood, at least qualitatively, within density 
functional theory. It came as a surprise, therefore, when several 
angle-resolved photoemission (ARPES)
studies on the intercalated layer compound Na$_{0.3}$CoO$_2$ \cite{ARPES}
revealed a fundamentally different shape of the Fermi surface than predicted 
by local density approximation (LDA) band theory \cite{singh}. 
Recent bulk sensitive Shubnikov-de Haas data \cite{SdH} also appear to 
be inconsistent with these calculations.
On the other hand, the overall width of the Co $3d$ bands in the ARPES
data was found to be only moderately reduced compared to the LDA value.
Essentially, the partially filled Co $3d$ $t_{2g}$ bands should give rise 
to a large $a_g$ hole pocket centered around $\Gamma$, and six small hole 
pockets of $e'_g$ character along the $\Gamma K$ directions of the hexagonal
Brillouin Zone. These $e'_g$ pockets have not yet been observed in 
experimental work. Their role for the superconducting hydrated phase  
of  Na$_{0.3}$CoO$_2$ is also a subject of intense investigations 
\cite{supercond}. In view of the narrow width of the Co $t_{2g}$ bands
($W\approx 1.5$~eV), one possible source of the discrepancy between ARPES
and band theory might be the effect of intra-$3d$ Co Coulomb interactions 
which, in principle, could enhance orbital polarization by leading to a 
charge transfer from $a_g$ to $e'_g$ subbands, and, eventually, to a 
shift of the $e'_g$ bands below the Fermi level.

The influence of Coulomb interactions on the topology of the Fermi 
surface of Na$_{0.3}$CoO$_2$ has been investigated by several groups, using
various theoretical methods and levels of approximation. Ishida {\it et al.}
\cite{ishida} applied dynamical mean field theory (DMFT) \cite{DMFT} based 
on the multi-orbital quantum Monte Carlo (QMC) method, together with a 
single-particle Hamiltonian derived from an accurate tight-binding fit 
of the $t_{2g}$ bands to the linearized augmented plane wave (LAPW) band 
structure. The result of this work was that, for Coulomb energies 
$U\approx3.0\ldots3.5$~eV and exchange $J=U/4$, the $e'_g$ hole pockets 
were slightly enlarged compared to the LDA Fermi surface, in contrast to 
the ARPES data. The width of the $t_{2g}$ bands, however, was found to be 
reduced to about 1~eV, in approximate agreement with ARPES. To avoid sign 
problems, only Ising-like exchange terms were included in the QMC calculation. 
It was also shown that an LDA+U \cite{LDAU} treatment can lead to enlarged 
or reduced $e'_g$ hole pockets, depending on whether $U$ is smaller or 
larger than $5J$, respectively.    

At the same time, Zhou {\it et al.}~\cite{zhou} investigated this problem 
within the Gutzwiller approach in the large $U$, $J=0$ limit. Using a
slightly different tight-binding fit to the LDA bands, these authors found
that the $e'_g$ bands were shifted below $E_F$, and that the width of the
$t_{2g}$ bands was strongly reduced from $1.5$~eV to about $0.5$~eV. Thus,
while the Fermi surface appears to agree with the ARPES data, the band
narrowing is much stronger than experimentally observed. Similar 
Gutzwiller calculations in the $U\rightarrow\infty$, $J=0$ limit were 
recently carried out by Shorikov {\it et al.}~\cite{shorikov}, with 
results similar to those of Ref.~\cite{zhou}. 
Since the Gutzwiller method replaces the frequency dependent complex 
self-energy by parameters providing orbital depending energy shifts and
band narrowing, it represents an approximation to DMFT. Moreover, for 
$U\rightarrow\infty$, complete orbital polarization is to be expected.
Thus, for a meaningful comparison with ARPES data, it is important  
to extend the Gutzwiller approach to realistic Coulomb and exchange 
energies appropriate for Co. 

The influence of correlations on the electronic properties of 
hydrated Na$_{0.35}$CoO$_2$ were also investigated by Landron and 
Lepetit~\cite{landron} within quantum chemical methods for embedded
CoO$_6$ and Co$_2$O$_{10}$ clusters. The crystal field splitting between
$a_g$ and $e'_g$ orbitals was found to be $\Delta=315$~meV, and the
Coulomb and exchange energies $U=4.1$~eV and $J=0.28$~eV. At present, 
it is not clear how these parameters, in particular, the large value 
of $\Delta$, would be modified for larger clusters that are required 
to describe the electronic properties of the extended system. Slave-boson 
mean field calculations by Bourgeois {\it et al.}~\cite{bourgeois} 
based on $\Delta=315$~meV and $U\rightarrow\infty$ revealed a 
pure $a_g$ Fermi surface and a $t_{2g}$ band width of $0.5$~eV, 
similar to the results of Ref.~\cite{zhou}. 

To examine the role of Hund exchange contributions not included in 
the QMC/DMFT calculations, Perroni {\it et al.}~\cite{perroni}  
applied a new multi-band exact diagonalization DMFT scheme to 
Na$_{0.3}$CoO$_2$. This approach does not suffer from sign problems
and includes density-density contributions as well as spin-flip and 
pair-exchange terms. Also, larger values of $U$ can be handled than 
via QMC. The result of this study was that in this material there is 
little difference between Hund and Ising exchange, and that, for 
$U=3\ldots5$~eV and $J=U/4$, the $e'_g$ pockets were slightly enlarged, 
in agreement with the QMC/DMFT results. Also, the band narrowing was 
found to be roughly 30~\%, consistent with the QMC 
treatment and with the ARPES data.  

Most recently, Marianetti {\it et al.}~\cite{marianetti} studied the 
problem of the $e'_g$ hole pockets in Na$_{0.3}$CoO$_2$ by applying a 
new continuous-time QMC/DMFT version that allows to reach larger $U$ 
and lower temperatures. The single-particle Hamiltonian was the same 
as in Ref.~\cite{zhou}, except for the crystal field splitting 
$\Delta=E_{a_g}-E_{e'_g}$ that shifts the $a_g$ bands up and the $e'_g$ 
bands down. For $U=3\ldots5$~eV and $J=0$, reduced $e'_g$ pockets are 
found for $\Delta=-10$~meV, and fully suppressed pockets if $\Delta$ is 
increased to $50\ldots 100$~meV. According to the authors, their results 
are ``in agreement with Ref.~\cite{zhou} and in disagreement with 
Ref.~\cite{ishida}''. Since the DMFT calculations, however, were not done 
for the same input Hamiltonian and $U$, $J$ values as those in 
Refs.~\cite{ishida,perroni}, the origin of the conflicting trends 
is presently unknown.   
 
The purpose of this work is to resolve this issue and to analyze
the role of the single-particle Hamiltonian $H(k)$ and Coulomb and 
exchange energies for the charge transfer between $t_{2g}$ bands. 
In particular, we show that the ED and QMC many-body calculations
are in perfect agreement if identical input parameters are employed.
On the other hand, the two different versions of $H(k)$ used in 
Refs.~\cite{ishida,perroni} and \cite{zhou,marianetti}
(below we refer to them as $H_1$ and $H_2$, respectively)
give rise to a slight, but significant difference in the variation 
of $e'_g$ occupancy with $U$: Whereas $H_1$ yields decreasing $e'_g$ 
occupancy with increasing $U$, $H_2$ gives the opposite trend. 
We show that these differences are caused by the $t_{2g}$ crystal 
fields $\Delta$ contained $H_1$ and $H_2$. The key point, however, 
is that, for realistic Coulomb and exchange energies, i.e., 
$U\approx 3\ldots 5$~eV and $J\approx 0.72$~eV~\cite{kroll}, 
the differences caused by $H(k)$ are small and do not affect the 
controversy concerning the shape of the Fermi surface. Both versions 
of $H(k)$ yield the result that, without an additional $a_g/e'_g$ 
crystal field splitting, Coulomb interactions do not eliminate the 
$e'_g$ hole pockets. The overall topology of the Fermi surface remains 
the same as predicted by LDA band theory. 

\begin{figure}[t]%1
  \begin{center}
  \includegraphics[width=3.8cm,height=8cm,angle=-90]{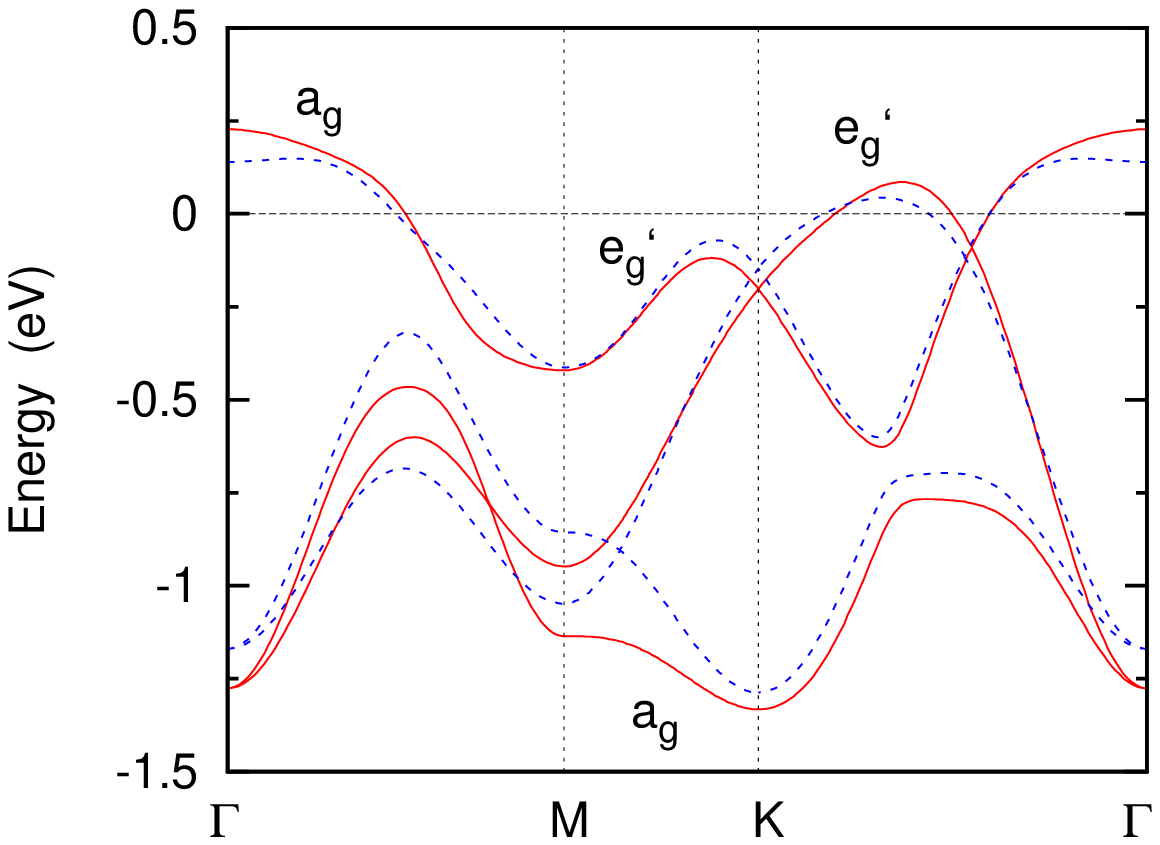}
  \includegraphics[width=3.8cm,height=8cm,angle=-90]{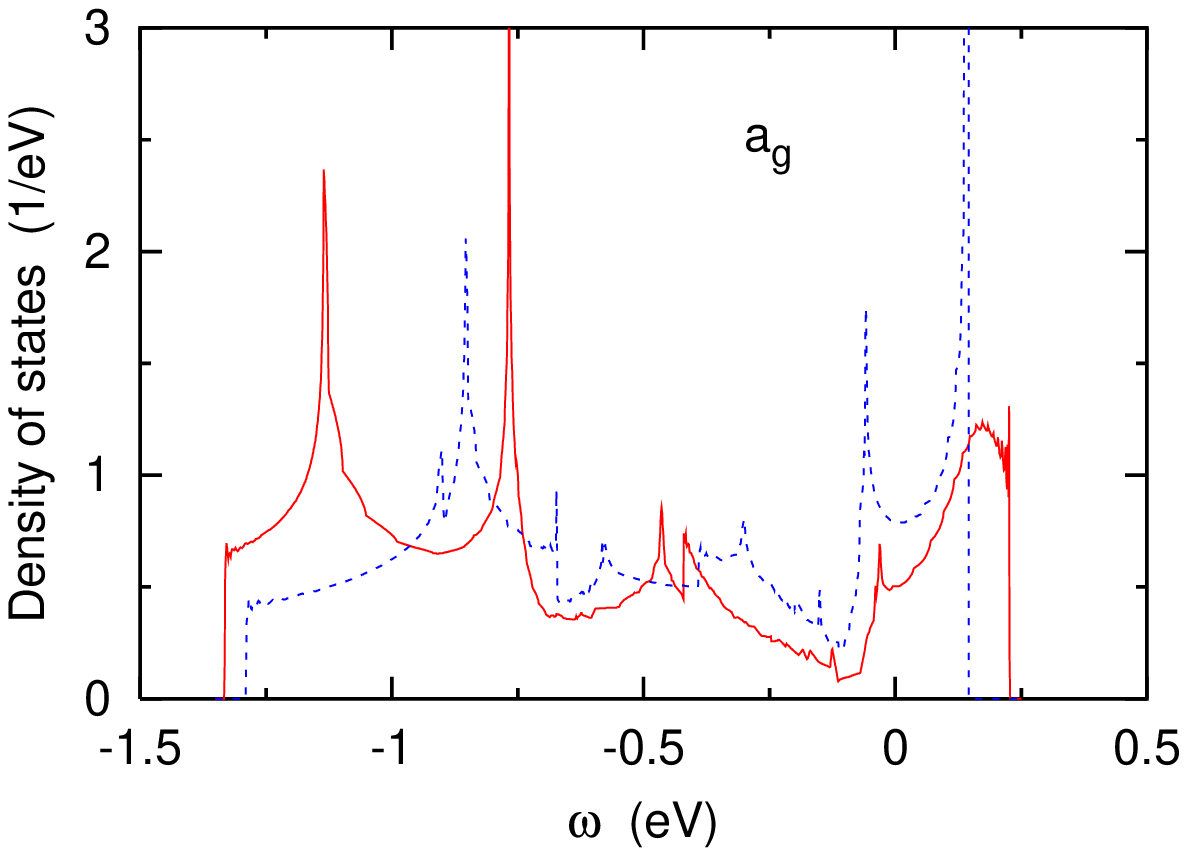}
  \includegraphics[width=3.8cm,height=8cm,angle=-90]{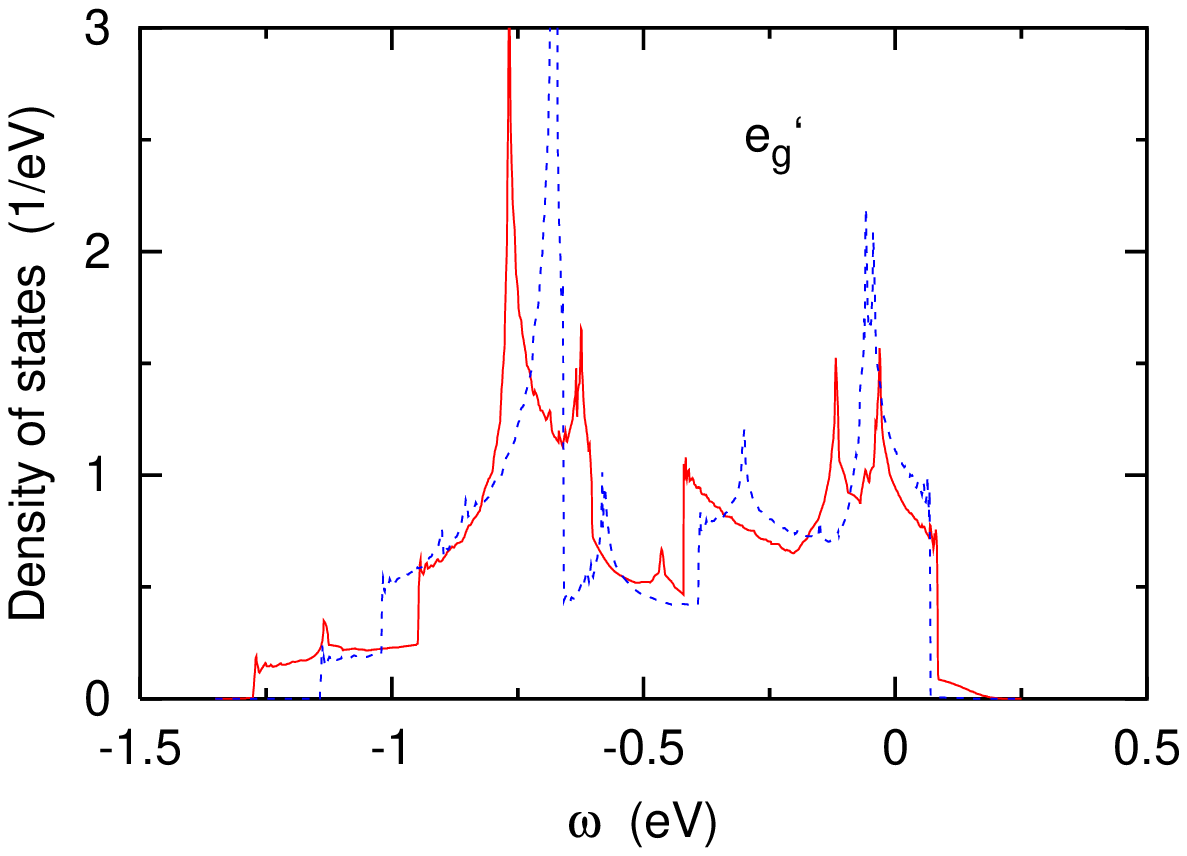}
  \end{center}
\vskip-3mm
\caption{
Upper panel:
Tight-binding fits to LDA $t_{2g}$ band structure of Na$_{0.3}$CoO$_2$
used in Refs.~\cite{ishida,perroni} and Refs.~\cite{zhou,marianetti}.
Lower panels: comparison of $a_{g}$ and $e'_{g}$ density of states. 
Solid (red) curves: $H_1$,  dashed (blue) curves: $H_2$, $E_F=0$.
}\end{figure}

Fig.~1 shows the tight-binding fits to the partially occupied Co 
$3d$ $t_{2g}$ bands used in Refs.~\cite{ishida,perroni} and 
\cite{zhou,marianetti}. The total occupancy is 5.3. 
The $a_g$ and $e'_g$ occupancies (per spin band) are 
$n_{a_g}\approx0.80$ and $n_{e'_g}\approx0.925$. Although both 
Hamiltonians give similar energy bands, they differ in a fundamental 
aspect: The predominant $a_g$ wave function character of the lowest 
LAPW band along $MK$~\cite{singh} is correctly reproduced via $H_1$, 
resulting in a van Hove singularity in the $a_g$ density of states 
near $-1.12$~eV. In contrast, the two lowest $H_2$ bands along $MK$
cross, so that this singularity is shifted to $-0.84$~eV, implying 
a significant effective narrowing of this subband. The lowest $H_2$ 
band at $M$ has $e'_g$ character and gives only a weak step at 
$-1.02$~eV in the density of states. Since the influence of Coulomb 
interactions is highly sensitive to the band width and the distribution 
of spectral weight within a band, these differences should affect also 
the correlation induced charge transfer between $t_{2g}$ bands. 

The top of the $H_2$ $a_g$ band at $\Gamma$ is seen to exhibit a 
minimum which is absent for $H_1$ and leads to a pronounced
peak in the density of states. This spectral weight is distributed 
over a slightly wider energy range in the case of $H_1$. The minimum 
is caused by interlayer interactions which are absent in $H_1$ as well 
as $H_2$. Moreover, the bulk $a_g$ density of states does not exhibit 
a sharp peak in this region. Thus, the $H_1$ $a_g$ density shown in 
Fig.~1 should be more appropriate than the one derived from $H_2$.   

Despite these differences, near $E_F$ both Hamiltonians yield 
nearly        identical bands. The $e'_g$ bands extend less than 100~meV 
above $E_F$, and both models exhibit the $a_g$\,/\,$e'_g$ crossing along 
$\Gamma K$ just below $E_F$. Note that these bands hybridize 
away from this symmetry direction, i.e., the crossing turns into an 
increasing hybridization gap as soon as the parallel momentum deviates 
from $\Gamma K$. 

We now discuss the correlation induced changes of the $t_{2g}$ bands
of Na$_{0.3}$CoO$_2$ as calculated within DMFT. We had previously 
demonstrated that, for the Hamiltonian $H_1$, the QMC
and ED results of Refs.~\cite{ishida,perroni} are in excellent 
agreement and that both schemes yield reduced orbital polarization
with increasing $U$. Moreover, this trend was found to be 
insensitive to the choice of $J$. 

%The $e'_g\rightarrow a_g$ charge 
%transfer was nearly the same for $J=0$ and $J=U/4$, with virtually
%no difference between anisotropic Ising and isotropic Hund exchange. 

We have applied the ED approach of Ref.~\cite{perroni} to $H_2$, 
in order to check its consistency with the QMC formalism used in 
Ref.~\cite{marianetti}. For $U=3$~eV, $J=0$, the subband
self-energies as a function of Matsubara frequency were found to be 
in almost quantitative agreement. In view of the inevitable slight 
numerical differences between these fundamentally different DMFT 
approaches, the excellent agreement between the ED and QMC results 
is indeed remarkable. We also point out that both DMFT schemes take 
proper account of static and dynamical correlations. 

The unexpected result of this calculation is that, with $H_2$ as 
input, both ED and QMC schemes yield {\it enhanced} orbital polarization: 
For $U=3$~eV, $J=0$, the subband occupancies are $n_{a_g}=0.735$, 
$n_{e'_g}=0.957$, compared to the LDA values $n_{a_g}=0.8$, 
$n_{e'_g}=0.925$. This charge transfer is opposite to the {\it reduced} 
orbital polarization obtained for $H_1$: $n_{a_g}=0.825$, $n_{e'_g}=0.91$.

Fig.~2 shows that similar systematic differences between $H_1$ and $H_2$ 
are found at other values of $U$, $J$. To eliminate other sources 
of possible differences, all results are derived using the ED/DMFT 
approach of Ref.~\cite{perroni} at $T=20$~meV. Regardless of the 
choice of $J$, $H_1$ leads to a reduction of $n_{e'_g}$ as a function 
of $U$, whereas $H_2$ yields increasing $n_{e'_g}$. Thus, although 
the single-particle band structure and density of states derived from 
$H_1$ and $H_2$ look qualitatively similar, these two Hamiltonians 
lead to a small, but significant difference in the variation of the 
subband occupancies with Coulomb energy.

To analyze the origin of this unusual behavior we simplify the evaluation 
of the quasi-particle Green's function
\begin{equation}
 G(i\omega_n) = \sum_{k}\, [i\omega_n + \mu - H(k) - \Sigma(i\omega_n)]^{-1},
\end{equation}
where $\omega_n=(2n+1)\pi/\beta$ are Matsubara frequencies, with 
$\beta=1/k_B T$ and temperature $T$. $G$, $H$ and $\Sigma$ are 
matrices in the $t_{2g}$ basis. Because of the planar 
hexagonal symmetry, the diagonal elements of $G$ are identical, and 
so are the off-diagonal elements. The same applies to $\Sigma$. 
In the $a_g$, $e'_g$ basis, these quantities become diagonal, 
with elements $G_a=G_{11}+2G_{12}$ and $G_e=G_{11}-G_{12}$, and
analogous expressions for $\Sigma_{a,e}$. 
In this basis, the Green's functions can be approximately written as 
\begin{equation}
G_i(i\omega_n) = \int\!\!d\omega \rho_i(\omega)
                 [i\omega_n + \mu - \omega - \Sigma_i(i\omega_n)]^{-1},
\end{equation}
where $\rho_i(\omega)$ are the  $a_g$ and $e'_g$ density of 
states components shown in Fig.~1. We have checked that Eq.~(2) yields
self-consistent DMFT solutions nearly identical to the ones derived from 
Eq.~(1). Thus, the different solutions obtained for $H_1$ and $H_2$ 
are directly related to the different shapes of the respective density
of states distributions. As is evident from Fig.~1, the $e'_g$ densities
are quite similar for both Hamiltonians. Indeed, replacing one by the
other does not alter the trends for the charge transfer 
shown in Fig.~2. It is clear, therefore, that the different shapes of 
the $a_g$ density of states are the source of the opposite orbital
polarization found for $H_1$ and $H_2$. 
 
\begin{figure}[t]%2
  \begin{center}
  \includegraphics[width=5.9cm,height=8cm,angle=-90]{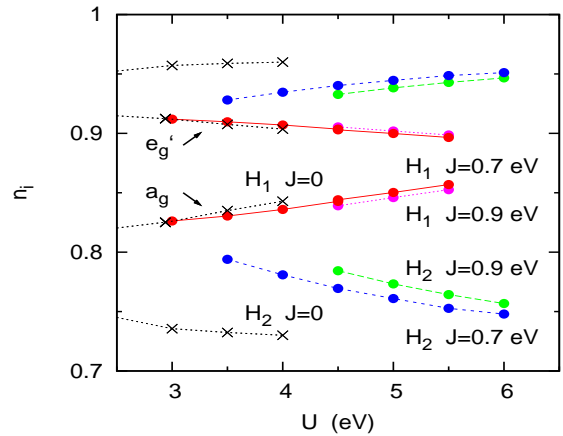}
  \end{center}
\vskip-3mm
\caption{
Subband occupancies as a function of $U$ for fixed $J$ derived 
within ED/DMFT for $T=20$~meV. 
Solid (red) curves: $H_1$, $J=0.7$~eV;
dotted (magenta) curves: $H_1$, $J=0.9$~eV;
dashed (blue) curves: $H_2$, $J=0.7$~eV;
long-dashed (green) curves: $H_2$, $J=0.9$~eV.
For comparison, some results for $J=0$ are also shown (crosses).
}\end{figure}
   
As pointed out above, the fact that the lowest $H_2$  bands
cross along $MK$ leads to an upward shift of the lowest $a_g$ van Hove 
singularity by about $0.3$~eV. One can simulate this redistribution
of spectral weight by reducing the $a_g$ density of $H_1$ in the range
$\omega<-0.7$~eV and amplifying it in the region $-0.7$\,eV$<\omega<0$, such
that the occupied weight remains $0.8$. This deformation is sufficient to
reverse the trend of $n_i(U)$ and give rise to a weak enhancement of 
orbital polarization. 

Evidently, the upward shift of spectral weight caused by the band crossing
along $MK$ implies a relative shift of $3d$ energy levels. In the case of 
$H_1$, the centroids of the $a_g$ and $e'_g$ density of states are 
$E_{a_g}=-0.624$~eV and $E_{e'_g}=-0.491$~eV, where $\Delta = E_{a_g} - 
E_{e'_g}=-133$~meV is the $t_{2g}$ crystal field splitting. In the 
tight-binding fit, $\Delta$ was varied along with the hopping parameters, 
in order to achieve the optimum representation of the LAPW bands throughout 
the Brillouin Zone~\cite{ishida}. Clearly, its value reflects the electronic 
structure of the extended system. In the case of $H_2$, the splitting was 
chosen as $\Delta =-10$~meV, and only the hopping parameters were 
fitted~\cite{zhou}. The $a_g$ and $e'_g$ centroids therefore nearly coincide:  
$E_{a_g}=-0.489$~eV and $E_{e'_g}=-0.479$~eV. 
Note that, in both cases, $E_{a_g}<E_{e'_g}$ despite $n_{a_g}<n_{e'_g}$. 

As a result of these different level splittings, correlations lead
to an intriguing reversal of interorbital charge transfer: For $H_1$ with
$\Delta =-133$~meV, the large $a_g/e'_g$ splitting is enhanced and gives 
rise to a gradual filling of the $a_g$ band with increasing $U$. Since, 
at small $U$, the $a_g$ occupancy is lower than the $e'_g$ occupancy, 
this charge transfer amounts to an initial reduction of orbital polarization. 
(At larger $U$, $n_{a_g}$ might become larger than $n_{e'_g}$, so that the
same correlation induced $e'_g\rightarrow a_g$ charge transfer eventually 
could turn into enhanced orbital polarization.) 
In contrast, the small crystal field included in $H_2$, $\Delta=-10$~meV,
is too weak to enforce a correlation induced downward shift of 
the $a_g$ band. Thus, the charge transfer is dominated by the larger 
$e'_g$ occupancy, giving enhanced orbital polarization even at small $U$.   

Since $H_1$ provides the more accurate fit to the LAPW bands, 
the correlation induced reduction of orbital polarization in the range 
of reasonable values of $U$ and $J$ should be more realistic than the 
opposite trend obtained for $H_2$.  Thus, as argued in  
Refs.~\cite{ishida,perroni}, correlations slightly enhance the $e'_g$ 
hole pockets of Na$_{0.3}$CoO$_2$. We emphasize, however, that, according 
to Fig.~2, the opposite charge transfer obtained for $H_2$ is not 
large enough to push the  $e'_g$ bands below $E_F$. 
                 
As pointed out in Ref.~\cite{marianetti}, a large positive crystal field 
can enhance orbital polarization, so that, in combination with local 
Coulomb interactions, the $e'_g$ hole pockets disappear~\cite{comment}.   
If we assume the ARPES data to be correct, the crucial question then 
concerns the physical origin of such a crystal field. Evidently, it 
is not related to on-site Coulomb interactions in the spirit of a 
single-site DMFT. Non-local effects stemming from the momentum 
dependence of the self-energy have not yet been explored and could 
be studied by using a cluster extension of the DMFT. Na disorder was 
shown to eliminate the pockets at large Na concentrations near 
$x=0.7$~\cite{singh2,maria2}, but is believed to be too weak to have a 
significant effect on the Fermi surface near $x=0.3$.  
Surface effects which have played an important role in ARPES data
on other perovskites, such as Ca$_{2-x}$Sr$_x$RuO$_4$ and 
Ca$_{1-x}$Sr$_x$VO$_3$, should also be investigated, in particular,
the effect of Na induced states on the first layer. Moreover, 
possible structural distortions, such as intra-planar buckling, and their 
connection to the opening of the $a_g/e'_g$ hybridization gap along 
$\Gamma K$ should be explored.

We finally mention that, as emphasized in Ref.~\cite{marianetti}, the 
$e'_g$ hole pockets are also important for the
understanding of the heat capacity of Na$_{0.3}$CoO$_2$. At present,
the experimental value, $\gamma\approx 12\ldots16$~mJ/molCoK$^2$, 
is difficult to reconcile with the LDA result, $\sim 14$~mJ/molCoK$^2$,
and an effective mass enhancement of about 2, as derived within the
DMFT studies discussed above. For a more detailed analysis of this 
quantity it might be necessary to relax the pinning condition implied
by the single-site approximation and allow for non-local effects.           

In summary, we have resolved the puzzling discrepancies between DMFT
results for the correlation induced $a_g/e'_g$ charge transfer in   
Na$_{0.3}$CoO$_2$. For identical input quantities, ED and QMC impurity
treatments are in excellent agreement. Surprisingly, however, slight 
differences among the 
tight-binding Hamiltonians lead to increasing or decreasing orbital 
polarization. This reversal of subband occupancies as a function of 
$U$ underlines the importance of using a high-quality single-particle
basis as input in the many-body calculation. In the present system, 
for realistic Coulomb and exchange energies, the differences resulting 
from these opposite trends are small, i.e., Coulomb interactions do not 
fill the $e'_g$ hole pockets. The topology of the Fermi surface of 
Na$_{0.3}$CoO$_2$ therefore remains the same as predicted by LDA band 
theory.
    
We hope that these results encourage further experimental work to study 
in more detail the geometrical and electronic structure of this
fascinating material.

One of us (A.L.) likes to thank Chris Marianetti for 
extensive correspondence and for sending his QMC self-energy results, 
and Michelle Johannes and Igor Mazin for very useful discussions.


\begin{thebibliography}{99}
\bibitem{comment}   
   $\Delta\approx 50\dots100$~meV for $J=0$.  For $J=0.7\ldots0.9$~eV,
   $\Delta$ should be even larger. It might then be necessary to readjust 
   the hopping parameters to describe the electronic properties 
   of the system characterized by a large $\Delta$.
\end{thebibliography}
\end{document}